\documentclass[prb,aps,floatfix,reprint,superscriptaddress]{revtex4-2}
\usepackage{amsmath}
\usepackage{amssymb}
\usepackage{amsfonts}
\usepackage{float}
\usepackage[colorlinks=true,citecolor=blue,linkcolor=magenta]{hyperref}
\usepackage{graphicx}
\usepackage{bm}
\usepackage{bbold}
\usepackage{color}
\usepackage[dvipsnames]{xcolor}
\usepackage{soul}
\usepackage{silence}
\DeclareMathAlphabet\mathbfcal{OMS}{cmsy}{b}{n}
\newcommand{\avg}[1]{{\langle#1\rangle}}

\newcommand{\ket}[1]{{|#1\rangle}}

\newcommand{\proj}[2]{{|#1\rangle\langle#2|}}
\newcommand{\inner}[2]{{\langle#1|#2\rangle}}
\newcommand{\tr}{{\mathrm{tr}}}
\newcommand{\im}{{\imath}}
\newcommand{\dg} {{\dagger}}
\newcommand{\pd} {{\phantom\dagger}}
\newcommand{\ai}[1] {{a_{#1}^{\pd}}}
\newcommand{\aid}[1] {a_{#1}^\dg}

\newcommand{\ci}[1] {{c_{#1}^{\pd}}}
\newcommand{\cid}[1] {c_{#1}^\dg}
\newcommand{\hc} {\text{h.c.}}
\newcommand{\bJ} {\bm{J}}
\newcommand{\bG} {\bm{G}}
\newcommand{\bS} {\bm{\Sigma}}
\newcommand{\bGa} {\bm{\Gamma}}
\newcommand{\bH}{\bm{\bar{H}}}
\newcommand{\bHS}{\bm{\bar{H}}_\qs}

\renewcommand{\Im}{\operatorname{Im}}
\newcommand{\ql}{\mathcal{L}}
\newcommand{\qs}{\mathcal{S}}
\newcommand{\qr}{\mathcal{R}}
\newcommand{\qi}{\mathcal{I}}

\newcommand{\qw}{\mathcal{W}}
\newcommand{\qd}{\mathcal{D}}
\newcommand{\NW}{N_\qw}
\newcommand{\NS}{N_\qs}
\newcommand{\cm}{\mathbfcal{C}}
\newcommand{\Zop}{\mathbfcal{Z}}
\newcommand{\Gop}{\mathbfcal{G}}
\newcommand{\Pop}{\mathbfcal{P}}
\newcommand{\Mop}{\mathbfcal{M}}
\newcommand{\DeltaF}{\Delta_F}
\newcommand{\DeltaS}{\Delta_\qs}
\newcommand{\barDeltaS}{\bar{\Delta}_\qs}
\newcommand{\Iav}{I_\diamondsuit}
\newcommand{\TotalS}{W_\qs}
\newcommand{\vS}{v_\qs}
\newcommand{\VS}{V_\qs}
\newcommand{\omegawidth}{\Gamma_{\omega}}
\newcommand{\conj}[1]{{#1}^{*}}

\newcommand{\Hsquare}{
   \text{\fboxsep=.0pt\fbox{\rule{0pt}{1ex}\rule{1ex}{0pt}}}
 }
\newcommand{\gammareference}{\gamma^{\Hsquare}}
\newcommand{\gammatransition}{\check{\gamma}}

\renewcommand{\[}{\begin{equation}}
\renewcommand{\]}{\end{equation}}

\begin{document}
\title{Transport in a periodically driven tilted lattice via the extended reservoir approach: Stability criterion for recovering the continuum limit}
\author{Bitan De}
\affiliation{Institute of Theoretical Physics, Jagiellonian University in Kraków, Łojasiewicza 11, 30-348 Kraków, Poland}
\author{Gabriela W\'ojtowicz}
\affiliation{Institute of Theoretical Physics, Jagiellonian University in Kraków, Łojasiewicza 11, 30-348 Kraków, Poland}
\affiliation{Biophysical and Biomedical Measurement Group, Microsystems and Nanotechnology Division, Physical Measurement Laboratory, National Institute of Standards and Technology, Gaithersburg, Maryland 20899, USA}
\affiliation{Doctoral School of Exact and Natural Sciences, Jagiellonian University in Kraków, Łojasiewicza 11, 30-348 Kraków, Poland}
\author{Jakub Zakrzewski}
\email{jakub.zakrzewski@uj.edu.pl}
\affiliation{Institute of Theoretical Physics, Jagiellonian University in Kraków, Łojasiewicza 11, 30-348 Kraków, Poland}
\affiliation{Mark Kac Center for Complex Systems Research, Jagiellonian University, ul. Łojasiewicza 11, 30-348 Kraków, Poland}
\author{Michael Zwolak}
\email{mpz@nist.gov}
\affiliation{Biophysical and Biomedical Measurement Group, Microsystems and Nanotechnology Division, Physical Measurement Laboratory, National Institute of Standards and Technology, Gaithersburg, Maryland 20899, USA}
\author{Marek M. Rams}
\email{marek.rams@uj.edu.pl}
\affiliation{Institute of Theoretical Physics, Jagiellonian University in Kraków, Łojasiewicza 11, 30-348 Kraków, Poland}
\affiliation{Mark Kac Center for Complex Systems Research, Jagiellonian University, ul. Łojasiewicza 11, 30-348 Kraków, Poland}

\begin{abstract}
Extended reservoirs provide a framework for capturing macroscopic, continuum environments, such as metallic electrodes driving a current through a nanoscale contact, impurity, or material. 
We examine the application of this approach to periodically driven systems, specifically in the context of quantum transport. 
As with non--equilibrium steady states in time--independent scenarios, the current displays a Kramers' turnover including the formation of a plateau region that captures the physical, continuum limit response. 
We demonstrate that a simple stability criteria identifies an appropriate relaxation rate to target this physical plateau. 
Using this approach, we study quantum transport through a periodically driven tilted lattice coupled to two metallic reservoirs held at a finite bias and  temperature. 
We use this model to benchmark the extended reservoir approach and assess the stability criteria. 
The approach recovers well--understood physical behavior in the limit of weak system--reservoir coupling. 
Extended reservoirs enable addressing strong coupling and non--linear response as well, where we analyze how transport responds to the dynamics inside the driven lattice. 
These results set the foundations for the use of extended reservoir approach for periodically driven, quantum systems, such as many--body Floquet states. 
\end{abstract}

\date{\today}
\maketitle

\section{Introduction}
\label{sec:introduction}
Quantum transport plays a central role in spectroscopy for many--body quantum systems, from superconducting and hybrid interfaces~\cite{wiedenmann_transport_2017,souto_multiterminal_2022,boolakee_light-field_2022} to quantum dot arrays~~\cite{wang_experimental_2022,roche_detection_2012,wang_spatially_2016,le_topological_2020,kiczynski_engineering_2022} to cold atoms~\cite{chien_interaction-induced_2013,gruss_energy-resolved_2018}. Transport can also serve as a probe of time--dependent states, such as time crystals within interacting, driven, dissipative quantum systems~\cite{sarkar_emergence_2022, sarkar_signatures_2022}. Yet, transport properties are challenging to compute and become even more so for time--dependent driving. 

We study the use of the extended reservoir approach (ERA) (see Ref.~\onlinecite{elenewski_communication_2017} for an overview) in obtaining transport characteristics for time--dependent transport. 
ERA is a rapidly developing area of research that employs a finite collection of reservoir modes to represent the continuum environment, including environments of many--body systems~\cite{arrigoni_nonequilibrium_2013, dorda_auxiliary_2014, dorda_auxiliary_2015, dorda_optimized_2017, chen_markovian_2019,fugger_nonequilibrium_2020,lotem_renormalized_2020,wojtowicz_open-system_2020, brenes_tensor-network_2020}.
To do so, ERA modes must be relaxed by external, implicit environments. 
However, one of the most useful flavors of ERA employs Markovian relaxation~\cite{gruss_landauers_2016}.
While being more computationally tractable, this relaxation breaks the fluctuation--dissipation theorem~\cite{kubo_fluctuation-dissipation_1966,gruss_landauers_2016,zwolak_analytic_2020}, which is only restored in an appropriate limit. 
The discreteness of the reservoirs also introduces anomalous virtual tunneling~\cite{wojtowicz_dual_2021}. 
These make the accurate calculation of transport properties a delicate limiting process, where one has to break (artificial) symmetries of the discrete model, identify an appropriate ({\em a moderate}) relaxation rate, and ensure the simulation is converging in a manner consistent with physical principles and continuum physics. 

We demonstrate how this process plays out for time--dependent systems.
These systems can introduce artificial resonances into the setup. 
As a model, we consider a periodically driven tilted lattice, the closed version of which is well studied in optical systems~\cite{yao_many-body_2020,chanda_coexistence_2020,lukin_probing_2019}. 
Without driving, such a lattice exhibits Wannier--Stark localization for large enough tilts~\cite{gluck_wannierstark_2002,fukuyama_tightly_1973,hartmann_dynamics_2004} {and transient Bloch oscillations~\cite{pinho_bloch_2023}. 
Introducing interactions  may result in rectification, as recently discussed for transport induced by Markovian reservoirs~\cite{mendoza-arenas_giant_2022}.}
Driving can lead to resonance--induced transport, or obstruct transport in other scenarios, in well--studied limiting cases~\cite{kohler_driven_2005}. 

When simulating this model with a finite collection of (relaxed) reservoir modes, i.e., the ERA approach, artificial resonant states can form across the reservoir--system--reservoir setup. 
These impart anomalous behavior to the current in a way that is {\em a priori} more difficult to recognize than for equivalent time--independent models. 
Avoiding this anomaly requires a larger relaxation rate than typical approaches employ or an additional averaging procedure. 
We show how a stability criteria identifies this rate and then use this approach to study the physical behavior of the driven, tilted lattice.

This paper is organized as follows: 
Section~\ref{sec:model} outlines the general transport framework, which can include both impurity and extended systems. 
We also introduce the periodically driven tilted lattice. 
Section~\ref{sec:ERA} summarizes the extended reservoir approach with Markovian relaxation, as well as presents and assesses the stability criteria to target a physical relaxation rate.
Section~\ref{sec:results} applies the approach to driven systems, as well as connects the results to closed systems for weak coupling, presents other validation procedures, and goes beyond linear response. 
We conclude in Sec.~\ref{sec:conclusions}.

\section{Transport framework and model}
\label{sec:model}

While much of what we develop is applicable to open quantum systems generally, such as those in the presence of a dissipative bosonic environment, we focus on quantum transport in this work and specifically on transport through a periodically driven fermionic system. 
In this section, we first introduce the transport framework and then the particular model we study. 

\subsection{Quantum transport}
\label{sec:transport}

The typical setup for transport has two macroscopic, i.e., continuum, reservoirs that connect to each side of a system. 
For time--independent scenarios, a finite bias (or temperature drop) across the reservoirs drives the system out of equilibrium and results in a current flow. 
Time--dependent systems can have richer behavior, as a, e.g., periodic drive can pump energy into the system and currents can flow even in the absence of an external bias. 

When the system (and system only in this work) can be time dependent, the general Hamiltonian is
\[ \label{eq:Ht}
H(t) = H_\qs(t) + H_\ql + H_\qr + H_{\qi}.
\]
The system $\qs$'s Hamiltonian, $H_\qs(t)$, is the region between the two reservoirs $\ql$ and $\qr$. In addition to having time dependence, this region generally will have many--body interactions, such as electron--electron or electron--vibration interactions. 
In this work, we will have only a quadratic system Hamiltonian in order to benchmark the ERA without additional complications. 
{The method we present ultimately aims at interacting models where the exact solution is not available. 
We can, however, develop a good understanding of ERA based on fully non--interacting models where exact reference solutions exist. 
Since we address issues with the reservoir representation, we expect our findings immediately generalize to many--body $\qs$ in contact with the same reservoirs.}

\begin{figure}[t]
    \includegraphics[width=\columnwidth, clip=true]{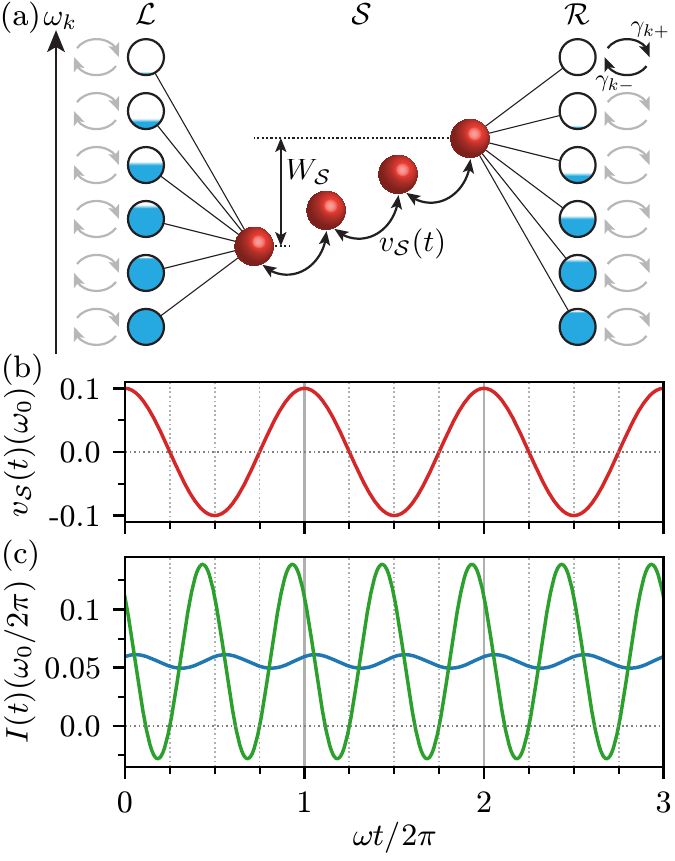}
    \caption{
    {\bf Quantum transport in the ERA.} 
    (a) A driven tilted lattice $\qs$ (red sites with total tilt $\TotalS$ and hopping $\vS(t)$) between two extended reservoirs, $\ql$ and $\qr$.
    The reservoirs are maintained at chemical potentials $\mu_\ql$ and $\mu_\qr$, represented by different filling (blue shading), and at the same temperature for the cases we examine.
    Discretized modes in $\ql$ and $\qr$ are explicitly included in the simulation.
    These modes are relaxed to their (isolated) equilibrium occupancy by  Markovian injection and depletion  at rates $\gamma_{k+}$ and $\gamma_{k-}$, respectively.
    (b) Periodically modulated hopping $\vS(t)$ inside system $\qs$ versus time. 
    (c) The induced particle current $I(t)$ versus time. 
    The blue line is the current from $\ql$ to $\qs$ and the green line between the middle sites in $\qs$.
    The simulations are for a $\NS=8$ site system driven at the resonant frequency $\omega=\DeltaS = \omega_0 / (N_\qs - 1)$, i.e., $\TotalS = \omega_0$. 
    The other parameters are $\VS = v_0 = \omega_0/10$, $\mu=\omega_0/2=2\mu_\ql=-2\mu_\qr$, and
    $k_B T_\ql = k_B T_\qr = \hbar \omega_0 /40$. The relaxation rates are set via the approach in Sec.~\ref{sec:ERA}.
    }
    \label{fig:schematic}
\end{figure}

Whether for many--body or non--interacting $\qs$, the two reservoirs are both continuum, non--interacting metallic reservoirs. The Hamiltonians are
\[ \label{eq:Halp}
H_{\alpha} = \sum_{k \in \alpha}  \omega_{k} \cid{k} \ci{k}
\]
for the $\alpha=\ql, \qr$ reservoirs, where $\omega_{k}$ is the frequency of the single--particle eigenstate $k \in \alpha$. 
We note that all Hamiltonians in this work are in terms of frequencies. 
The last contribution, $H_\qi$, is the interaction between $\qs$ and $\ql\qr$, which we take to be quadratic hopping only,
\[ \label{eq:HI}
H_{\qi} = \sum_{k \in \ql \qr} \sum_{i \in \qs} \left( v_{ki} \cid{k} \ci{i} + v_{ik} \cid{i} \ci{k} \right),
\]
with $v_{ki}=\conj{v}_{ik}$. This $H_{\qi}$ is the typical paradigm whether $\qs$ is non--interacting or many body. The $\cid{m}$ ($\ci{m}$) are the fermionic creation (annihilation) operators for mode $m$.
The index $m$ carries all necessary mode labels, such as frequency, spin, and region ($\ql$, $\qs$, or $\qr$). 
We will use $k$'s ($l$'s) and $i$'s ($j$'s) to indicate single--particle eigenstates of $\ql\qr$ and spatial modes of $\qs$, respectively. 

For non--interacting reservoirs coupled linearly to the system with the number conserving interaction in Eq.~\eqref{eq:HI}, the behavior of the setup is determined by the reservoirs' spectral functions,
\[ \label{eq:Jl}
\bJ_\alpha(\omega) = 2 \pi \sum_{k \in \alpha} \proj{v_k}{v_k} \delta(\omega - \omega_k),
\]
for $\alpha = \ql, \qr$.
The $\bJ_\alpha(\omega)$ is a square matrix of size equal to the number of sites in the system $\qs$.
To keep the notation compact, we use a coupling vector $\ket{v_k}$ between mode $k\in\ql\qr$ and all sites $i\in\qs$, i.e., $\inner{i}{v_k}=v_{ik}$. 
The general aim of ERA is to recover the macroscopic limit via a finite number of broadened reservoir modes within the reservoir bandwidth $\qw$. 
These modes must capture all relevant features encoded in the continuum $\bJ_\alpha(\omega)$, as well as how that spectral density is populated according to the Fermi-Dirac distribution. 

We are most interested in the particle current in a Floquet state that has a periodic drive in the presence of a bias in the reservoirs' chemical potentials $\mu_\ql$ and $\mu_\qr$. When taken at the $\ql$ and $\qs$ interface, this current is
\[ \label{eq:curr}
    I_{\ql \qs}(t) = 2 \Im \sum_{k \in \ql} \sum_{j \in \qs} v_{jk} \avg{\cid{j} \ci{k}}_t,
\]
where $\avg{\cdot}_t$ indicates the quantum mechanical average at time $t$.
The current has a similar form at the other interfaces, all following from continuity equations.
{Since the Hamiltonian in Eq.~\eqref{eq:Ht} conserves total particle number, the current follows from considering time dependence of local occupations induced by $H_{\qi}$ in Eq.~\eqref{eq:HI}.}
While the time dependence of the current can be different for the various interfaces, we focus on the mean current. 
For periodic driving, the average need only be over a single oscillation period $\tau = 2 \pi / \omega$,
\[ \label{eq:Iav}
    \Iav =\frac{1}{\tau} \int_{0}^{\tau} I_{\ql\qs}(t) dt,
\]
in the Floquet state.
This quantity is interface--independent in the physical limit of interest.

In all our examples, we consider a uniform, low temperature of $k_B T_\ql = k_B T_\qr = \hbar \omega_0 /40$, where $k_B$ is Boltzmann's constant and $\hbar$ the reduced Planck's constant. 
A temperature bias could also be present, but we do not consider that case.
We take a symmetrically applied potential $\mu$, i.e., $\mu_\ql = - \mu_\qr = \mu/2$, and the hopping strength, $\omega_0$, as a reference frequency. 
For most examples, this is the actual hopping in the reservoirs, which are uniform one--dimensional lattices. 
We, however, treat these reservoirs numerically in their single--particle eigenbasis, e.g., Eq.~\eqref{eq:Halp}, and their spatial dimensionality is not of central importance. 
We break the correspondence between $\omega_0$ and the hopping only where indicated in order to further validate the ERA via the fully Markovian limit.

\subsection{periodically driven tilted lattice}
\label{sec:tilted_lattice}

We consider a one--dimensional tilted lattice with $\NS$ sites and nearest--neighbor hopping, as schematically depicted (within the ERA) in Fig.~\ref{fig:schematic}(a). The time--dependent Hamiltonian is 
\[ \label{eq:HS}
H_\qs(t) = \sum_{i=1}^{\NS} \omega_{i}(t) \cid{i} \ci{i} + \sum_{i=1}^{\NS-1} \vS(t) (\cid{i} \ci{i+1} + \cid{i+1} \ci{i}).
\]
The tilt is linearly increasing with nearest--neighbor step $\DeltaS(t)$ inside $\qs$, giving the on--site frequencies
\[ \label{eq:tilt}
    \omega_i(t) = \DeltaS(t) \left( i - \frac{\NS + 1}{2} \right),
\]
for $i = 1, 2, \ldots, \NS$.
{This tilt is symmetric around zero frequency, following a similar choice for the reservoirs below.
In all our examples, apart from Sec.~\ref{sec:holthaus}, we consider a static tilt $\DeltaS(t)=\DeltaS$, with the total tilt
\begin{equation}
    \label{eq:DeltaS}
    \TotalS = \DeltaS (\NS - 1),
\end{equation}
and a hopping that oscillates as}
\begin{equation}
    \label{eq:vs}
    \vS(t) = \VS \cos(\omega t),
\end{equation}
with frequency $\omega$ and amplitude $\VS$. 
Figure~\ref{fig:schematic}(b) shows the oscillation of $v_\qs(t)$ along with the resulting oscillating currents (across two interfaces) in Fig.~\ref{fig:schematic}(c). 

As already indicated, we consider $\ql$ and $\qr$ to be uniform one--dimensional (semi--infinite) lattices coupled, respectively, to the first and last site of $\qs$ with hopping $v_0$ (i.e., $v_{ik} \neq 0$ only for $k\in \ql$, $i = 1$ and $k\in \qr$, $i = \NS$).
The Fermi level is at zero. 
This gives the continuum limit spectral functions 
\[ \label{eq:J1d}
    [\bJ_\ql(\omega)]_{11} = [\bJ_\qr(\omega)]_{\NS \NS} = \frac{8 v_0^2}{\qw^2} \sqrt{\qw^2 - 4 \omega^2},
\]
and zero otherwise. In terms of the reference frequency, $\omega_0$, the bandwidth is $\qw = 4 \omega_0$ unless otherwise indicated. 
Those exceptions will be calculations to further validate the ERA approach by showing how it converges to the fully Markovian limit.

The driven tilted lattice model provides a benchmark example for the ERA. 
It is well--studied in the context of closed systems, has non--trivial behavior versus the driving frequency, and allows for extended systems $\qs$. 
{Without driving}, $v_\qs(t) = \VS$, the system is a standard tilted lattice leading to Wannier--Stark localization~\cite{gluck_wannierstark_2002,fukuyama_tightly_1973,hartmann_dynamics_2004}.
In this limit, one expects efficient transport for a small global tilt $\TotalS$ and  suppression of transport when the global tilt exceeds the width of the Bloch band,
\[ \label{eq:delocalization}
|\TotalS| \gtrsim 4 |V_\qs|. 
\]
This condition becomes strict for large $\NS$.
At a non--zero driving frequency, the driven system allows for effective mapping to a static setup when the coupling to the reservoirs is weak, providing additional validation. 
The ERA, however, is not limited to those special cases and can be employed, for instance, also in the limit of strong or moderate system--reservoir couplings.

Moreover, this model, including the coupling to reservoirs, may be realized experimentally with a slight modification of the approaches described in Ref.~\cite{krinner_two-terminal_2017}. 
Recently, similar tilted models (with interactions) gained significant attention in the studies of many-body localization without disorder~\cite{schulz_stark_2019, van_nieuwenburg_bloch_2019, taylor_experimental_2020, chanda_coexistence_2020, yao_many-body_2020, scherg_observing_2021, guo_stark_2021, morong_observation_2021, yao_nonergodic_2021, yao_many-body_2021}.
Here, we consider the non--interacting lattice, however, to validate the ERA and set the foundations to simulating interacting cases with tensor networks~\cite{wojtowicz_open-system_2020}.

\section{Extended reservoir approach}
\label{sec:ERA}

The exact solution for the transport problem of Sec.~\ref{sec:transport} is in the macroscopic limit, where the reservoirs are a continuum and have an infinite--dimensional Hilbert space.
To make the problem tractable, we employ the {\it extended reservoir approach} (ERA)~\cite{gruss_landauers_2016,elenewski_communication_2017}.
In ERA, the reservoirs are approximated by a finite collection of explicit modes/sites, which, in turn, are coupled to implicit reservoirs. 
The latter relax the explicit modes to an (isolated) equilibrium state to maintain set temperatures and chemical potentials. There is a long lineage of relaxation--based approaches, starting from early work of Kohn and Luttinger~\cite{kohn_quantum_1957} to open--system approaches for semiconductors~\cite{frensley_simulation_1985,frensley_boundary_1990,knezevic_time-dependent_2013} to approximate master equations~\cite{sanchez_molecular_2006,subotnik_nonequilibrium_2009}. 
The presence of implicit relaxation supports a stationary state and, within ERA, provides a limiting process to capture the influence of continuum reservoirs on transport~\cite{gruss_landauers_2016,elenewski_communication_2017}. 
In this section, we explain the concept in detail, including both the discretization of the reservoirs and the relaxation, and introduce a stability criterion to set the main parameter of the computational approach.

\subsection{Discretization}
\label{sec:discretization}

A standard strategy to capture the influence of the reservoirs on the  system is to include them directly in the many-body calculation.
For numerical simulation, this requires discretization of the continuum, approximating it by a finite collection of $\NW$ modes.
In principle, any microscopic discretization that limits to the desired macroscopic spectral function as $\NW \to \infty$ works. 
This leaves considerable freedom. 
For instance, one can employ a lattice mapping~\cite{chin_exact_2010,binder_reaction_2018} or an inhomogeneous placement of modes, such as lin--log (evenly distributed inside the bias window and logarithmically outside)~\cite{schwarz_lindblad-driven_2016} or influence--based (leading to mode density vanishing as inverse frequency squared outside of the bias window)~\cite{zwolak_finite_2008,elenewski_performance_2021} distributions.
At finite bias, inhomogeneous distributions will, at best, give a prefactor improvement in the number of $\NW$ required, as the bias window requires a uniform distribution of modes (at most, one could exploit spectral function structure in the bias window).
For tensor networks, where the entanglement in the setup dictates the numerical cost rather than the bare number of modes, they may give no speedup at all~\cite{elenewski_performance_2021}.

We work with semi--infinite, uniform one--dimensional reservoirs. 
To obtain the discretized lattice for numerical simulation, we  truncate this reservoir to a lattice of $\NW$ sites. 
All simulations here employ these modes in the single--particle eigenbasis given by a sine transform, 
{
\[ \label{eq:1dwk}
    \omega_k = \frac{\qw}{2} \cos\left( \frac{k \pi}{\NW + 1} \right) + \delta_{\alpha}, 
\]
where $k = 1,\ldots, \NW$ and $\delta_\alpha$ is a small perturbation of the discretization which globally shifts the energies in reservoir $\alpha = \ql, \qr$ to which $k$ belongs.
}
Since reservoirs are attached to terminal sites of the system $\qs$, $v_{k1}=v_k$ for $k \in \ql$ and $v_{k\NS}=v_k$ for $k \in \qr$, with 
\[ \label{eq:1dvk} 
    v_k = v_0 \sqrt{\frac{2}{\NW + 1}} \sin \left( \frac{k \pi}{\NW + 1} \right) 
\]
and the other $v_{kj}$ in Eq.~\eqref{eq:HI} are equal to zero.

There is an additional class of parameters in Eq.~\eqref{eq:1dwk} above, a set of small frequency shifts, $\delta_\ql$ and $\delta_\qr$.
As described in the subsequent sections, we will use them as control parameters to probe the stability of the results and guide the selection of simulation parameters. 
The shifts are of the order of the level spacing,
\begin{equation}
    \label{eq:deltaF}
    \DeltaF \simeq  \frac{\pi \qw}{2 \NW},
\end{equation}
in the discretized reservoir at the Fermi level.

Now, one may initialize the setup in the desired initial state, generate a particle imbalance between reservoirs, and run the Hamiltonian evolution.
However, such an approach can only support a quasi--steady state at intermediate times limited by the size of the finite reservoirs.
That can make some protocols or parameter regimes hard to access.
Among others, even in the time--independent setup, the Gibbs phenomena related to the finite reservoir bandwidth can result in oscillating currents in the quasi-steady state~\cite{zwolak_communication_2018}.
For periodic driving, the latter could interfere with the driving frequency, making the continuum limit even harder to extract.

\subsection{Open--system approach}
\label{sec:open}
To address such challenges, a growing number of methods augment the explicit reservoirs with a relaxation process.
These go beyond closed--system approaches, such as the microcanonical approach~\cite{ventra_transport_2004,bushong_approach_2005,sai_microscopic_2007,chien_bosonic_2012,chien_landauer_2014}, and explicitly relax reservoir modes to their equilibrium distributions at the desired temperatures and chemical potentials.
Most approaches to date employ continuous relaxation within a Markovian master equation.
This has been done for classical thermal transport~\cite{velizhanin_driving_2011, chien_tunable_2013,velizhanin_crossover_2015,chien_thermal_2017, chien_topological_2018}, for non--interacting electrons~\cite{sanchez_molecular_2006,subotnik_nonequilibrium_2009,dzhioev_super-fermion_2011,ajisaka_nonequlibrium_2012,ajisaka_nonequilibrium_2013,zelovich_state_2014, zelovich_moleculelead_2015, schwarz_lindblad-driven_2016, zelovich_driven_2016, hod_driven_2016, elenewski_communication_2017,zelovich_parameter-free_2017, chiang_quantum_2020}, including for time--dependent driving~\cite{chen_simple_2014, oz_numerical_2020, lacerda_quantum_2023, brenes_particle_2022}, but also for interacting systems utilizing tensor network techniques in the simulations~\cite{arrigoni_nonequilibrium_2013, dorda_auxiliary_2014, dorda_auxiliary_2015, dorda_optimized_2017, chen_markovian_2019,fugger_nonequilibrium_2020,lotem_renormalized_2020,wojtowicz_open-system_2020, brenes_tensor-network_2020}.
This builds on the original concept of pseudo-modes~\cite{imamoglu_stochastic_1994,garraway_decay_1997,garraway_nonperturbative_1997,zwolak_dynamics_2008}, where external Markovian relaxation broadens the modes into Lorentzian peaks and turns a discrete reservoir into an effective continuum.
Different relaxation schemes can also be used, such as the recently introduced periodic refresh~\cite{purkayastha_periodically_2021}, which stroboscopically resets the reservoirs to their (isolated) thermal equilibrium, or a generalization that interpolates between periodic and continuous relaxations via the accumulative reservoir construction~\cite{wojtowicz_accumulative_2023}.

The stroboscopic refresh processes have advantages, potentially allowing for algebraically faster convergence versus $\NW$ of physical quantities to the continuum limit~\cite{wojtowicz_accumulative_2023}. 
However, in this work, we focus on a continuous Markovian treatment where the density matrix of $\ql\qs\qr$ follows a Lindblad master equation,
\[ \label{eq:evolution_cr}
    \dot{\rho} = -\im [H, \rho] + \qd [ \rho ],
\]
where the dissipative term is
\begin{eqnarray}
    \label{eq:dissipator}
    \qd [ \rho ] & = & \sum_{k\in\ql\qr} \gamma_k^+ \left( \cid{k} \rho \ci{k}
    - \frac{1}{2} \left\{ \ci{k} \cid{k} , \rho \right\} \right) \\
    && + \sum_{k\in\ql\qr} \gamma_k^- \left( \ci{k} \rho \cid{k}
    - \frac{1}{2} \left\{ \cid{k} \ci{k} \rho \right\} \right).
\end{eqnarray}
The $\left\{ \cdot, \rho \right\}$ gives the anticommutator with the density matrix $\rho$. 
The injection rates $\gamma_k^+=\gamma f^{\alpha_k}(\omega_k)$, and depletion rates, $\gamma_k^-=\gamma (1-f^{\alpha_k}(\omega_k))$, with free parameter $\gamma$, are such that the reservoirs in isolation relax to the thermal equilibrium defined by the Fermi-Dirac distribution
\[ \label{eq:fd}    
    f^{\alpha}(\omega)=\frac{1}{1+e^{\beta_\alpha(\omega-\mu_\alpha)}},
\]
with the thermal relaxation time $\beta_\alpha = \hbar/k_B T_\alpha$.
The relaxation rate $\gamma$ is the central control parameter and needs to be tuned to best mimic the continuum reservoirs. 
We elaborate on this in the next section.

For non--interacting systems, Eq.~\eqref{eq:evolution_cr} is  efficiently solvable for a Floquet state utilizing standard correlation matrix techniques (see Appendix~\ref{sec:cm}). 
The Lindblad form of the master equation, Eq.~\eqref{eq:evolution_cr}, in principle, allows direct treatment of a general, interacting system, where the density matrix can be conveniently approximated as a matrix product state~\cite{zwolak_mixed-state_2004,verstraete_matrix_2004}. 
However, turning matrix product states into a useful approach to tackle quantum transport requires careful selection of the computational basis and its ordering~\cite{rams_breaking_2020,wojtowicz_open-system_2020} to avoid exponential entanglement barrier precluding successful matrix product state simulations.
We focus on non--interacting systems, leaving the determination of the optimal structure of matrix product state simulations for time--dependent situations to future studies.

\begin{figure}[t!]%
    \includegraphics[width=\columnwidth]{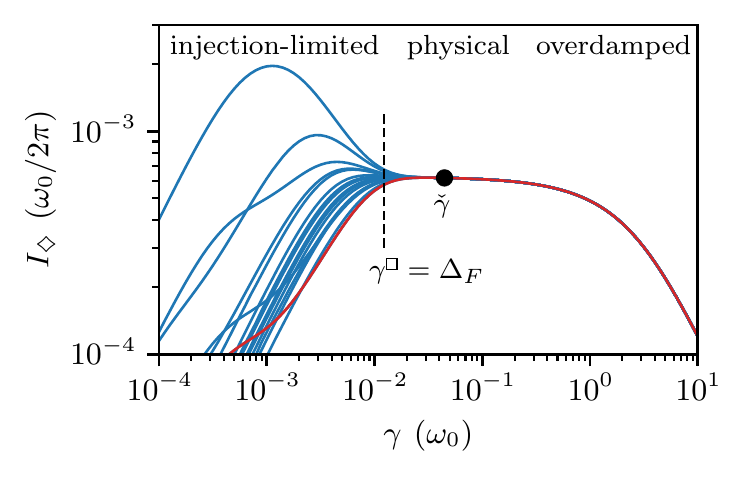}
    \caption{
    {\bf Stability and discreteness.}
    We plot Kramers' turnover for $\Iav(\gamma)$, indicating the approximate division into injection--limited, physical, and overdamped regimes. 
    The solid lines are different combinations of small discretization shifts in Eq.~\eqref{eq:shift}, $s_r,s_m \in \{ 0, \frac13, \frac23, 1\}$. The red line singles out a standard choice of $s_m=s_r=0$, i.e., no shift. 
    For small enough $\gamma$, resonances between discrete modes artificially modify the current.
    The black circle shows the transition relaxation rate, $\gammatransition$, that marks the end of the stable regime, i.e., the physical plateau, for a given reservoir size, $N_{\qw}=512$.
    The dashed line marks the standard choice of $\gamma^\Hsquare = \Delta_F$, where the current noticeably varies with the shift.
    The data are for $N_{\qs}=2$, a weak reservoir--system coupling $v_0=\omega_0 / 100$, $\DeltaS=\omega=\omega_0 \pi /16$, $V_{\qs}=\omega_0 / 20$, $\qw=4\omega_0$, and bias $\mu \to \infty$. 
    }
    \label{fig:2}%
\end{figure}%
\subsection{Stability criterion and the relaxation rate}
\label{sec:stability}

Green's function techniques permit a formal proof~\cite{gruss_landauers_2016} (for time--independent, non--interacting, and interacting systems) that the steady state of Eq.~\eqref{eq:evolution_cr} converges to the continuum limit with the current given by the Meir-Wingreen formula~\cite{meir_landauer_1992, jauho_time-dependent_1994} (e.g., for non--interacting systems, it converges to the Landauer formula ). 
{For time--dependent non--interacting models, one can find the solution using time--dependent non--equilibrium Green's functions~\cite{stefanucci_time-dependent_2008,moskalets_scattering_2011,gaury_numerical_2014}, which are exact up to the truncation of the frequency expansion.} 
We recover this limit for ERA by first taking $\NW \to \infty$ and then $\gamma \to 0$. 
In practical simulations, one simultaneously increases $\NW$ while decreasing $\gamma$, but still at quite modest $\NW$.

A standard choice is to set $\gamma$ proportional, and typically equal, to the level spacing $\DeltaF$ in Eq.~\eqref{eq:deltaF}, which should allow a sum of the discrete Lorentzians to approximately reproduce a desired spectral function (see, e.g., Refs.~\cite{chen_simple_2014,schwarz_lindblad-driven_2016,brenes_tensor-network_2020, lacerda_quantum_2023,brenes_particle_2022}). 
As well, one can approximate $\gamma$ from the self-energies of explicit reservoir modes in contact with the implicit infinite environment~\cite{zelovich_parameter-free_2017}, which for the reservoirs here would give $\gamma \approx 1.7 \Delta_F$. 
Such reasoning, however, considers the reservoir in isolation from the rest of the setup, which we will show here can be poorly behaved (although even for static models, virtual resonances can dominate the current depending on transmission properties of the system~\cite{wojtowicz_dual_2021}). 
We will use
\[ \label{eq:gammasquare}   
    \gammareference = \Delta_F
\]
as a reference relaxation rate and demonstrate that this choice can lead to systematic errors due to the presence of $\qs$, and thus always requires further validation. 
We also note that one often makes $\gamma$ mode dependent, but this would have marginal influence on our results. 
We employ a homogeneous $\gamma$ here for simplicity.

\begin{figure}[t!]%
    \includegraphics[width=\columnwidth]{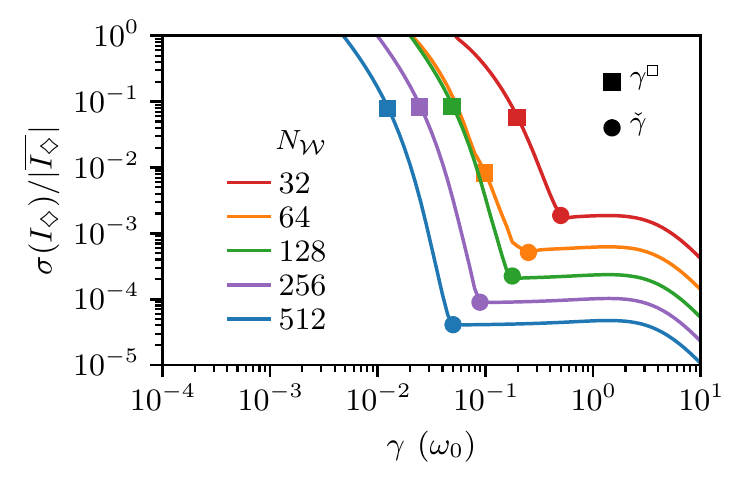}
    \caption{
    {\bf Stable and unstable relaxation.} 
    We show the maximal deviation of the current from the median current, normalized by the median, calculated from a combination of small discretization shifts in Eq.~\eqref{eq:shift}: $s_r,s_m \in \{ 0, \frac13, \frac23, 1\}$ (i.e.,~all lines in Fig.~\ref{fig:2}).
    The circles indicate $\gammatransition$, which marks the transition between a stable regime with deviation proportional to the shift, and an unstable one for smaller $\gamma$'s.
    The squares represent a standard choice of $\gammareference = \Delta_F$, where the relative precision (the sensitivity to a small discretization shift) does not improve with $\NW$.
    The results are for $N_{\qw}=32,64,128,256,512$ and other parameters as in Fig.~\ref{fig:2}.
    }
    \label{fig:3}%
\end{figure}

It is illustrative to treat $\gamma$ as a free parameter and consider its influence on the steady state.
For a fixed $\NW$, three basic regimes appear for the steady--state current, forming a so--called Kramers'  turnover~\cite{velizhanin_crossover_2015}.
The main regime of interest is a physical regime at intermediate $\gamma$ where the current is independent of $\gamma$ to leading order. 
The current is thus only weakly distorted by the reservoir approximation and it approaches the physical value of interest.
This plateau regime is flanked by large--$\gamma$ and small--$\gamma$ regimes.
In those overdamped and injection--limited regimes, respectively, the steady--state current is dominated by the relaxation rate and vanishes algebraically with that rate. 
This behavior is clearly visible in Fig.~\ref{fig:2} for the periodically driven system.
Such a dependence of the current on the relaxation rate mimics Kramers' turnover for chemical reaction rates~\cite{kramers_brownian_1940}.

A more detailed analysis reveals additional anomalous regimes that may appear at both ends of the physical regime, shifting its precise boundaries. 
In particular, on the low--$\gamma$ end of the physical plateau, a resonance between the discrete reservoir modes in   $\ql$ and $\qr$ can result in virtual transitions, artificially enhancing the current and masking the physical result~\cite{wojtowicz_dual_2021}.
Other conditions may also put constraints on the parameter ranges sufficient to recover physically relevant results. 
For instance, the $\gamma$--related broadening should be much smaller than thermal broadening, i.e., $\gamma\ll k_B T / \hbar$~\cite{elenewski_communication_2017}, at the same time this provides a lower bound to the required $\NW$. 

The above considerations have been extensively studied within time--independent setups, but should naturally generalize to time--dependent $\qs$ since they are properties of the reservoirs (albeit, the system can impact whether anomalous features are visible in the Kramers' turnover). 
Time dependence of $\qs$ only adds to the richness of phenomena. 
This motivates the data--driven approach here for estimating the relaxation rate that best reproduces the physical characteristics of transport.

We probe the stability of the current (or other properties) to small perturbations of the reservoir mode placement. 
While other options are possible, we employ small frequency shifts in Eq.~\eqref{eq:1dwk}, 
\begin{eqnarray}
    \label{eq:shift}
    \delta_{\ql} & = & (s_m / 2 + s_r / 4) \DeltaF, \\ \notag
    \delta_{\qr} & = & (s_m / 2 - s_r / 4) \DeltaF,
\end{eqnarray}
where the parameter $s_r \in [0, 1]$ controls the relative shift between $\omega_k$'s in $\ql$ and $\qr${, and $\DeltaF$, Eq.~\eqref{eq:deltaF}, is the mode spacing at the Fermi level}.
The $s_r$ influence artifacts coming from resonances between discrete $\ql$ and $\qr$ modes.
A mutual shift of both reservoirs with respect to $\qs$ is controlled by $s_m \in [0, 1]$.

For given discretization and relaxation rate, 
our stability test scans over a set of $(s_m, s_r)$ values to determine the robustness of the results against the shifts.
For larger $\gamma$ (see Fig.~\ref{fig:2}), small shifts do not affect the current since discrete modes are strongly broadened by coupling to implicit reservoirs.
With decreasing $\gamma$, the Lorentzian broadening of $\omega_k$ gets smaller, eventually revealing discrete nature of the reservoirs as seen by the system.
In this case, the overlap of the broadened levels (in both the reservoirs and system) becomes sensitive to small shifts,  leading to potential artificial features in the current.

\begin{figure*}[t!]%
    \includegraphics[width=2.00\columnwidth, clip=true]{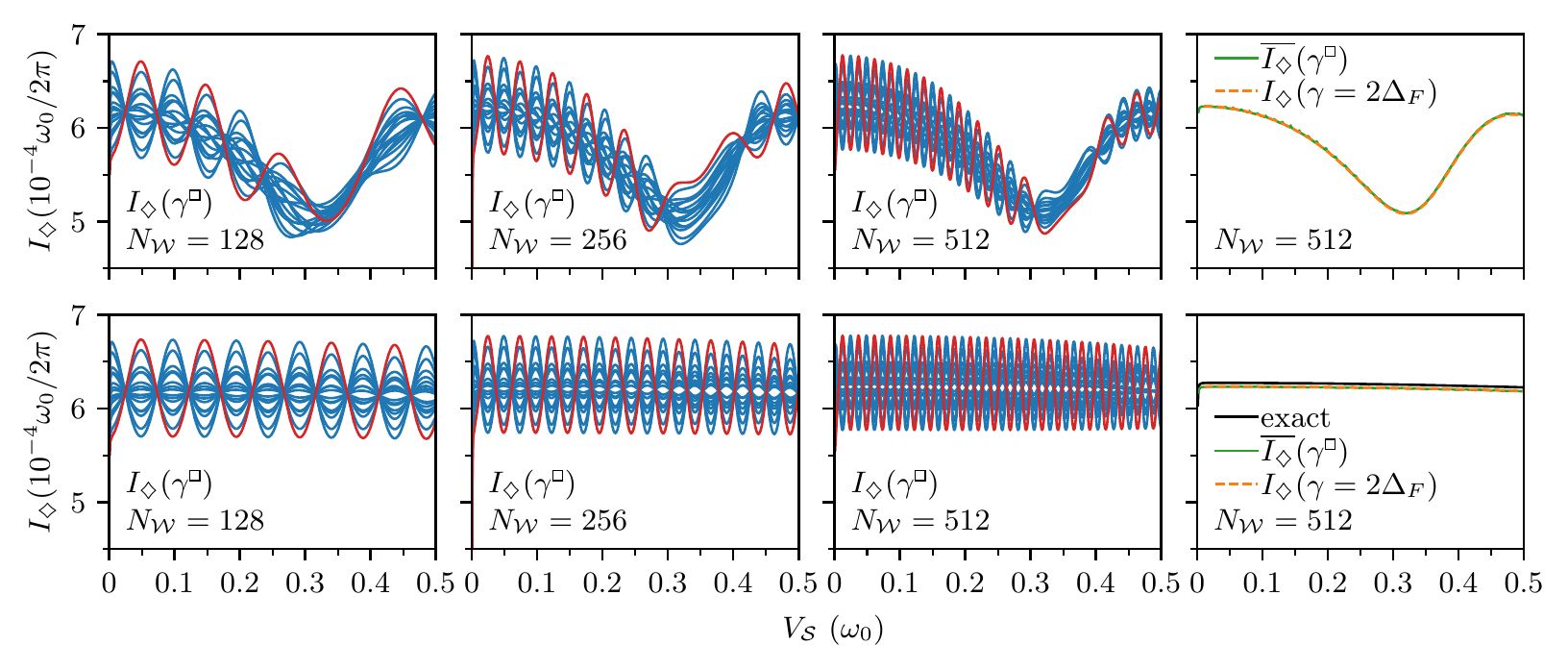}
    \caption{
    {\bf Failure of the standard relaxation choice.}
    Setup is the same as in Fig.~\ref{fig:2}, but we scan the coupling amplitude $\VS$ in $\qs$. Top row represents time--dependent simulations, and the bottom row shows the results within RWA. {In the first three columns,} the red line is a result of an often used $\gammareference = \Delta_F$ (without the shift, $s_m=s_r=0$), with blue lines showing deviation on small discretization perturbation with $s_m, s_r \in \{0, \frac13, \frac23, 1\}$ for the same $\gammareference${, calculated for growing number of reservoir modes $\NW$}. {In the last column, we show the median current over those shifts (green lines), and a current at a stable $\gamma = 2\Delta_F$ (orange lines). Those two results closely overlap each other. The proper convergence is further corroborated in RWA by comparing with the thermodynamic limit results (Landauer formula; black line).} The instability of the results for {standard $\gammareference$ without shifts} is of a similar order to {actual physical} features in the {sufficiently converged} $\Iav(\VS)$ {in the last column}, and does not appreciably decrease with growing $\NW$. 
    }
    \label{fig:S1}%
\end{figure*}%

A systematic scan for a range of $\NW$'s is in Fig.~\ref{fig:3}, where we plot the normalized maximal deviation from the median current, calculated over a set of $(s_m, s_r)$ values. 
It exhibits a clear transition between a stable regime for larger $\gamma$, where the small influence is proportional to the amplitude of the shift $\DeltaF$, and a non--linear regime for smaller $\gamma$.
The relaxation rate, $\gammatransition$, is at the transition between these regimes, marked with circles in Fig.~\ref{fig:3}.
Still, in practical simulations, one may choose a smaller $\gamma$, provided the precision given by the stability test is satisfactory.
In Fig.~\ref{fig:3}, the standard choice of $\gamma$ fails the stability test, giving a systematic error that does not decrease with increasing $\NW$. 
Figure~\ref{fig:S1} shows an additional example of such a failure where systematic errors distort the scan over system $\qs$’s coupling parameter persisting for increasing $\NW$. 
The current is represented more accurately by the median current at $\gammareference$ and by the current at
higher $\gamma$ where the instabilities are sufficiently removed.

The stability test can be used while scanning different values of $\gamma$ to help identify the range corresponding to the physical plateau (for better corroboration of a proper convergence to the continuum limit).
It can also be used without a full scan of $\Iav(\gamma)$ to estimate the precision at a particular $\gamma$ choice.
The criterion is both intuitively appealing and mathematically necessary: So long as the reservoirs limit to the same continuum spectral function, they describe the same model.
Thus, if they are providing different results (diverging curves in Fig.~\ref{fig:2}), then they do not represent  continuum reservoirs.

The stability criterion naturally generalizes the procedure of  Refs.~\cite{elenewski_performance_2021,wojtowicz_dual_2021}, which addressed virtual transitions between discretized reservoirs' modes as the source of instability.
In that procedure, we considered two turnovers ($s_r=0,1$ with $s_m=0$) and estimated the appropriate relaxation as the crossing point between them (i.e., the smallest deviation).
The extension proposed here is numerically more comprehensive, recognizing artificial effects from all possible resonances. 
Those might be particularly hard to identify prior to the calculation for time--dependent models, as well as in interacting or otherwise complicated systems.

\section{Results}
\label{sec:results}
We now implement this framework to study the periodically driven tilted lattice in Sec.~\ref{sec:tilted_lattice}.
We benchmark our results in the parameter limits where reference solutions are available and also cover cases where the simplifying approximations are no longer valid.

\subsection{Markovian limit}
\label{sec:vs_markonian}
In general, the evolution of $\qs$ is inherently non--Markovian when coupled to reservoirs.
However, it can become Markovian in some limits. 
One such case is the limit of infinite bandwidth and infinite bias, i.e., $\qw \to \infty$ with $\ql$ fully occupied and $\qr$ empty~\cite{elenewski_communication_2017}, where the evolution of $\qs$ is governed by the Markovian master equation
\begin{equation}
    \label{eq:markov_limit}
    \dot{\rho}_\qs = -\im\left[ H_\qs, \rho_\qs \right] + \qd_\qs [ \rho_\qs ]
\end{equation}
with the dissipative term
\begin{eqnarray}
    \qd_\qs [ \rho_\qs ] & = & \gamma^+_{\ql} \left( \cid{1} \rho_\qs \ci{1} - \frac{1}{2} \left\{ \ci{1} \cid{1} , \rho_\qs \right\}  \right) \\
    && + \gamma^-_{\qr} \left( \ci{N_\qs} \rho_\qs \cid{N_\qs} - \frac{1}{2} \left\{ \cid{N_\qs} \ci{N_\qs} , \rho_\qs \right\} \right),
\end{eqnarray}
which describes particle injection at the first system site and depletion from the last site{~\cite{landi_nonequilibrium_2022}}.

\begin{figure}[t!]%
    \includegraphics[width=0.99\columnwidth, clip=true]{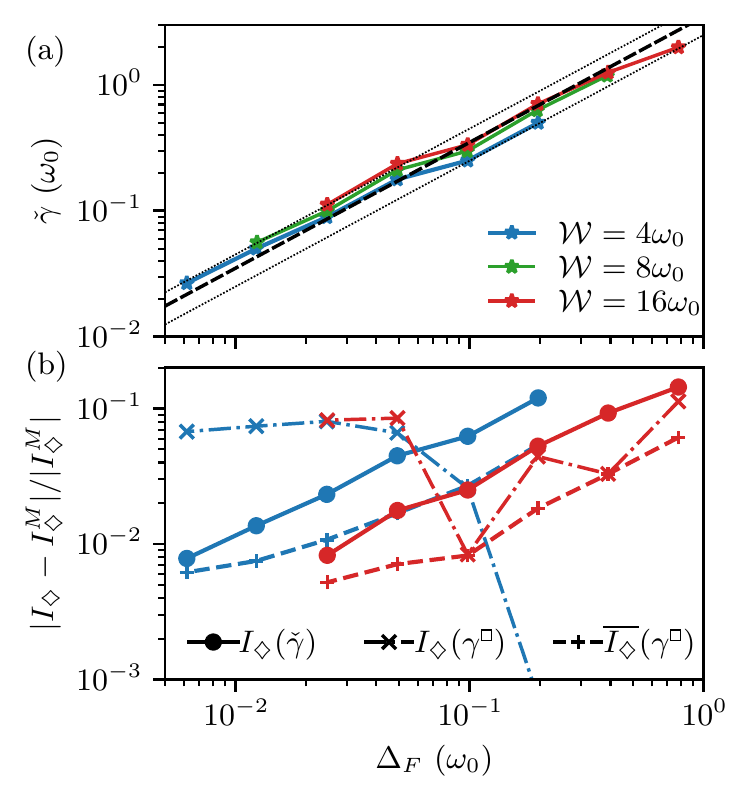}
    \caption{
    {\bf Convergence to the infinite bandwidth limit.}
    (a) The $\gammatransition$ from the stability criterion,  Sec.~\ref{sec:stability}, versus {$\DeltaF\propto\qw/\NW$}, Eq.~\eqref{eq:deltaF}, for $\NW = 32, 128, \ldots, 1024$. 
    Each $\gammatransition$ comes from the stability of Kramers' turnover to small perturbation of reservoir discretization, as in Fig.~\ref{fig:3}. 
    For small enough $\DeltaF$, $\gammatransition \simeq A \DeltaF$ (dashed line).
    The fit proportionality coefficient is $A = 3.5 \pm 1.0$, 
    {where the error indicates the maximal uncertainty range demarcated by the dotted lines.}
    (b) The relative deviation from the Markovian result (infinite bandwidth and infinite bias) versus $\DeltaF$. 
    Circles show the current at $\gammatransition$, where ERA systematically converges to the infinite bandwidth result. 
    Crosses show the results at $\gammareference$ (with no shift in the discretization), which gives rise to systematic errors with increasing $\NW$.
    {Finally, with pluses, we indicate a median over various shifts calculated at $\gammareference$, where the systematic errors are removed, and the results systematically converge. For clarity of the plot, we only show $\qw=4\omega_0$ (blue) and $\qw=16\omega_0$ (red), with $\qw=8\omega_0$ giving qualitatively the same results.}   Other parameters are as in Fig.~\ref{fig:3}.}
    \label{fig:4}%
\end{figure}%

We use it to approximate a setup with finite--bandwidth reservoirs, characterized by the spectral functions in Eq.~\eqref{eq:J1d}, and bias $\mu \to \infty$.
In that case, the injection and depletion rates are 
\begin{equation}
    \label{eq:gammaLR}
    \gamma^{+}_{\ql} = \gamma^{-}_{\qr} = 8 v_0^2 / \qw,
\end{equation}
{which follows from Born-Markov approximation for weak coupling $v_0$.}
Finite $\qw$ will give corrections to the Markovian approximation, which, for $\qw$ much larger than other energy scales, vanish with an extra factor of $\qw^{-1}$, {see Appendix~\ref{sec:markovian} for the derivation.}

In Fig.~\ref{fig:4}, we compare the results of the Markovian approximation in Eq.~\eqref{eq:markov_limit} to the outcome of the ERA.
First, in Fig.~\ref{fig:4}(a), we show the transition $\gammatransition$ for a series of $\NW$ and bandwidths. 
While it is proportional to the reservoir level spacing $\DeltaF$ to a good approximation, the proportionality coefficient takes a value $3.5 \pm 1.0$ in this example. 
In Fig.~\ref{fig:4}(b), we compare the resulting currents. 
The ERA results at $\gammatransition$ systematically converge with increasing $\NW$, approaching the expected Markovian limit with growing $\qw$. 
The results calculated at $\gammareference$ (with no shifts) are systematically shifted from the expected value, which results in the error saturating as $\NW$ grows. {We can, however, converge to the true current at $\gammareference$ by taking a median over various shifts (note that this is not inconsistent with Fig.~(3), which shows a maximal perturbation--related error). The convergence is less regular than at $\gammatransition$ but has a similar overall rate. Yet, the error is smaller overall for the median estimate at a given $\NW$, which is due to the fact that the corrections to the current from the Markovian anomaly are smaller at $\gammareference$ since it is a weaker relaxation (i.e., there is less distortion of the broadened modes but still a sufficient relaxation to look continuum like). 
}

\begin{figure}[t!]%
    \includegraphics[width=\columnwidth]{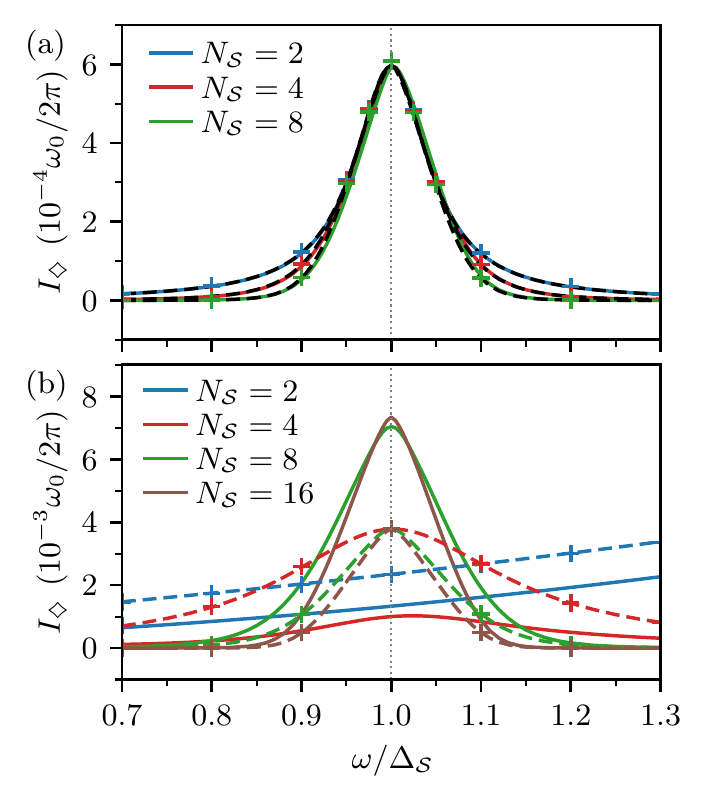}
    \caption{
    {\bf Resonant current through a tilted lattice.}
    (a) Current versus driving frequency showing the resonance for weak coupling to the reservoirs, $v_0= \omega_0 / 100$. 
    Time--dependent ERA results for $\NS=2,4,8$ (solid lines) and time--independent RWA results (dashed lines for thermodynamic limit and symbols for ERA) follow each other closely.
    The resonance width is consistent with  Eq.~\eqref{eq:sigma_omega}, reflected by the observed collapse of curves for various $\NS$.
    (b) Current versus driving for strong coupling, $v_0=\omega_0$, where the RWA provides a less accurate approximation of the time--dependent results.  
    The RWA captures the behavior of the resonance width (for larger $\NS$) but poorly quantifies the current amplitude. 
    In both panels, the inter--site tunneling, $\VS = \omega_0 / 20$, is weak and smaller than the driving frequency (the latter follows from the amplitude of the total tilt, $\TotalS = \omega_0$). Typically, one expects RWA to be a reasonable approximation of $\qs$ in this limit.
    We keep the bias at zero $\mu=0$, and the ERA results are obtained for $\NW=512$ and $\gamma=2 \Delta_F$. We checked all points for shift--related stability, with maximal relative deviation below $0.004$ (apart from time--dependent simulations for $v_0=\omega_0$ and $\NS=4$, where it is below $0.02$).    
    A small deviation between RWA results in the continuum limit and ERA, visible at the resonance in (a), is dominated by finite--$\NW$ error and can be reduced by increasing $\NW$.
    }
    \label{fig:5}%
\end{figure}%

This shows that one should backup a typical choice of $\gammareference$ with further tests, such as assessing stability and/or scanning $\gamma$ to identify the extent of a physical plateau in a particular model. These tests can be local or semi--local (stability at a fixed value of $\gamma$ or varying $\gamma$ in the vicinity of $\gammareference$) and still provide an estimate of precision.

\subsection{Rotating--wave approximation}
\label{sec:rwa}

The fully Markovian approach in Eq.~\eqref{eq:markov_limit} is quite simple to implement and is thus frequently used in the literature.
However, it cannot describe finite potential or a temperature bias, nor can it capture nontrivial reservoir features and their interplay with system dynamics.
In this section, we contrast ERA results with those from a different common approximation.

Let us revisit our system $\qs$, a tilted lattice given by Eq.~\eqref{eq:HS}. 
Recall first the time--independent system with $v_\qs(t) = \VS$. 
With a tilt, a single--particle problem can yield Stark localization.
Coupling the system to reservoirs with a reasonably weak coupling does not affect it, leading to an inhibition of transport for strong enough total tilt $\TotalS$ [see Eq.~\eqref{eq:delocalization}].

Periodically driven tunneling $v_\qs(t)$ destroys localization when the driving frequency $\omega$ is close to the frequency difference $\DeltaS$.
This can be seen from Eq.~\eqref{eq:HS} in an interaction picture.
{One can use new operators,}
\begin{equation}
    \label{eq:aj}
    \ai{j} \equiv e^{\im \bar{w}_j t} \ci{j},
\end{equation}
that are a unitary rotation $\exp ( \im t \sum_{k\in\ql\qs\qr} \bar{w}_k \cid{k} \ci{k} )$ of the old. 
In the rotating frame, the original Hamiltonian with the oscillating coupling $v_\qs(t)$ in Eq.~\eqref{eq:vs} becomes
\begin{eqnarray}
    \label{eq:HSI}
    H_\qs (t) &=&  \sum_{j=1}^{\NS} \left(w_j - \bar{w}_j \right) \aid{j} \ai{j} \\
    && + \frac{\VS}{2} \sum_{j=1}^{\NS-1} \left[\aid{j} \ai{j+1} (1+e^{-2 \im \omega t}) + \hc \right],
\end{eqnarray}
{when $\bar{w}_j = \omega \left( j - (\NS + 1) / 2 \right)$ for $j \in \qs$, compare with Eq.~\eqref{eq:tilt}. 
The gap between system sites becomes $\barDeltaS = \DeltaS - \omega$, which vanishes when $\DeltaS = \omega$ (i.e., in resonance). 
In the rotating wave approximation (RWA), one neglects the fast rotating terms $e^{\pm2 \im \omega t}$ in the above Hamiltonian, making the approximate model time--independent.}

One also needs to rotate the reservoirs to keep the $\qs$ and $\ql \qr$ coupling time independent. {
By the coupling of $\ql$ to left--most site of $\qs$ the $\ql$'s modes accumulate a shift
$ {\omega}_{k} \rightarrow  {\omega}_{k} + \omega (\NS - 1) / 2$. Similarly, the $\qr$ reservoir's modes 
shift to $ {\omega}_{k} \rightarrow  {\omega}_{k} - \omega (\NS - 1) / 2$.
This effectively moves the $\ql$ and $\qr$ bandwidths out of alignment, and the chemical potentials follow as $ {\mu}_{\ql} \rightarrow  {\mu}_{\ql} + \omega (\NS - 1) / 2$ and $ {\mu}_{\qr} \rightarrow  {\mu}_{\qr} - \omega (\NS - 1) / 2$. 
Consequently, in the rotated frame, the bias appears as
\begin{equation}
    \label{eq:muRWA}
    \bar{\mu} = \mu + \omega \TotalS / \DeltaS,
\end{equation} 
i.e., there is an effective bias due to the drive.
}

In Fig.~\ref{fig:5}, we compare the time-averaged ERA solution with the time--independent RWA predictions.
The latter permits using the Landauer formula, valid for non--interacting time--independent setups, to obtain results directly in the limit of continuum reservoirs (see Appendix~\ref{sec:landauer}). We also present ERA results applied to RWA Hamiltonian for further corroboration.

The RWA is expected to hold near resonance for driving that is much faster than other scales in the system, in particular for weak $V_\qs$ and $v_0$.
In Fig.~\ref{fig:5}(a), we show the results for a weak coupling to reservoirs, $v_0 = 0.01 \omega_0$, where we can expect RWA to work extremely well. 
We observe a full agreement between the two approaches.
We may use RWA to estimate the width of the resonance in Fig.~\ref{fig:5}.
Combining the localization/delocalization condition in Eq.~\eqref{eq:delocalization} and RWA Hamiltonian following from Eq.~\eqref{eq:HSI}, the resonance peak width, $\omegawidth$, should satisfy $\omegawidth (\NS - 1) \sim \VS$. {Using Eq.~\eqref{eq:DeltaS}, i.e., that $\TotalS=\DeltaS (\NS - 1)$, this} translates to
\begin{equation}
    \label{eq:sigma_omega}
    \frac{\omegawidth}{\DeltaS} \sim \frac{4 \VS}{\TotalS}.
\end{equation}
Indeed, the data collapse in Fig.~\ref{fig:5}(a) for all $\NS$ since all curves have the same $\VS / \TotalS$, {and we plot the current as a function of $\omega/\DeltaS$}. One expects RWA to hold around the peak of the resonance, {where $\omega/\DeltaS = 1$}.

In Fig.~\ref{fig:5}(b), we show the data for strong coupling to the reservoirs, $v_0=\omega_0$. 
Here, the RWA approximation is no longer valid as the strong coupling to the reservoirs leads to a fast transport through the system and, at such short times, the averaging of the counter--rotating $\exp(\pm 2\im\omega t)$ terms is less effective.

\begin{figure*}[t!]%
    \includegraphics[width=2\columnwidth]{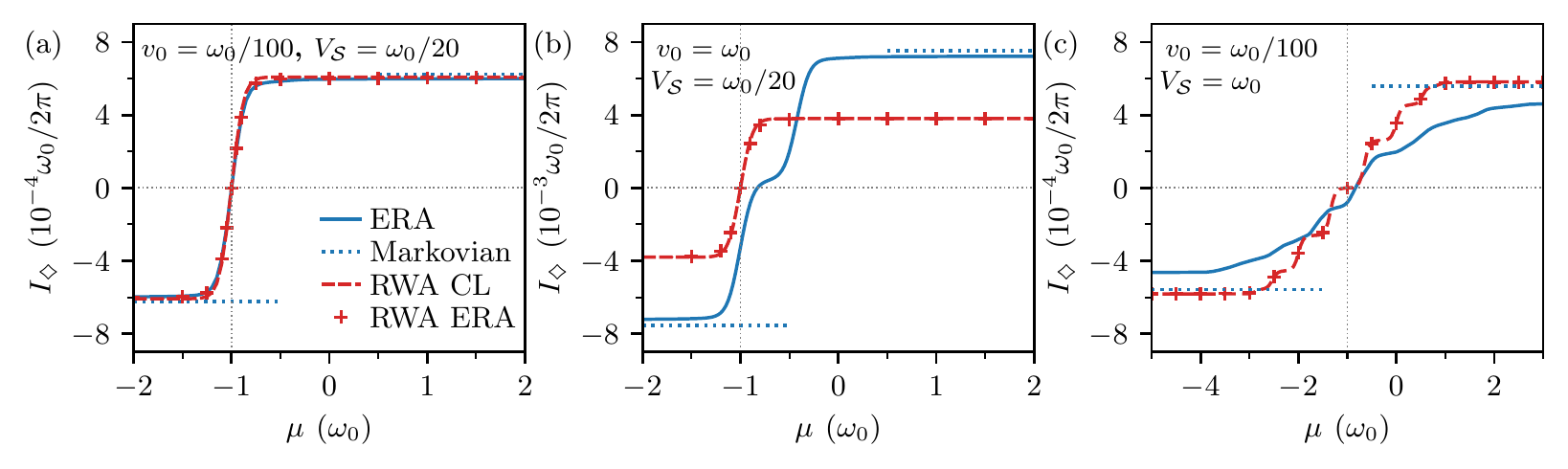}
    \caption{
    {\bf Current versus bias at resonance.}
    We consider various combinations of the system--reservoir coupling and hopping amplitude in $\qs$: in (a), $v_0 = \omega_0 / 100$ and $\VS = \omega_0 / 20$; in (b), $v_0=\omega_0$ and $\VS = \omega_0 / 20$; in (c), $v_0=\omega_0 / 100$ and $\VS = \omega_0$.
    We compare the ERA results with both RWA and Markovian approximation results of Eq.~\eqref{eq:markov_limit}. 
    The latter captures the current relatively well in the limit of large biases.
    It is, however, unable to witness the zero crossing when the influence of periodic driving compensates for applied bias.
    Approximate results for the RWA are directly in the continuum limit (CL) (dashed lines) and using ERA simulations (symbols).
    They closely match the time--dependent simulations allowed by ERA (solid lines) in (a), with zero crossing for $\mu = -\TotalS = -\omega_0$.
    For strong $v_0$, in (b), RWA provides a rudimentary picture where the actual current, captured by ERA, has a different amplitude and shifted position of the zero crossing. 
    Increasing $\VS$ in (c), the increased level spacing in finite $\qs$ leads to current quantization within time--independent RWA, which, unlike in (a) and (b), is not washed out here by a small but finite reservoir temperature $T_\ql = T_\qr = \hbar \omega_0 / 40 k_B$.
    Such steps get smoothed out in the actual periodically-driven setup.
    Data for a system of $\NS=8$ sites, total tilt $\TotalS=\omega_0$, and the bandwidth $\qw=4 \omega_0$. 
    The ERA results are obtained for $\NW=1024$ and $\gamma=0.01$ that we checked for stability. 
    }
    \label{fig:6}%
\end{figure*}%

In Fig.~\ref{fig:6}, we focus on the resonance, $\omega = \DeltaS$, and show the current as a function of the potential bias $\mu$. 
In this case, the effective bias in the rotated frame is $\bar{\mu} = \mu + \TotalS$.
Consequently, RWA predicts that periodic driving of $v_\qs(t)$ leads to a non--zero current even when the applied bias is zero, $\mu=0$, and that the direction of the current changes, crossing zero for $\mu = - \TotalS$.
Figure \ref{fig:6}(a) shows the case of weak coupling to the reservoirs when the system dynamics is dominant.
The ERA approach is able to correctly recover the Fermi level in the reservoirs, and the threshold of the current precisely matches the RWA prediction.
For a stronger coupling $v_0$, presented in Fig.~\ref{fig:6}(b), this simple picture breaks down for the reasons explained already. 
The precise position of the zero crossing gets noticeably shifted from RWA prediction of $\mu = - \TotalS$. 
Also, the amplitude of the current gets underestimated by RWA.
The Markovian approximation of Eq.~\eqref{eq:markov_limit} is better at recovering the current amplitude in the limit of large (negative or positive) bias.
However, by its very nature, it is unable to describe the effective compensation of a finite bias by periodically driving.

Finally, in Fig.~\ref{fig:6}(c), we keep the coupling to reservoirs weak and increase $\VS$.
This allows witnessing the presence of discrete energy levels in a finite system $\qs$.
In the RWA, $\qs$ forms a finite lattice without a tilt, translating to eigenfrequencies $V_\qs \cos \left(l \pi/(\NS+1)\right)$ with $l=1,2,\ldots,\NS$ (analogously to Eq.~\eqref{eq:1dwk}). 
Effectively, each level contributes to transport when it lies within the bias window controlled by $\bar{\mu}$.
This results in visible steps in the current for the RWA,  Fig.~\ref{fig:6}(c), as changing $\mu$ includes successive system eigenenergies in the bias window.
Such steps are smoothed out for weak $V_\qs$ in Figs.~\ref{fig:6}(a) and \ref{fig:6}(b) due to thermal broadening (note that we fix $T_\ql = T_\qr = \hbar \omega_0 /40 k_B$).
Similarly, this explains the saturation of currents in Fig.~\ref{fig:6} for sufficiently large bias when all transition channels in $\qs$ participate in transport.
As discussed above, the RWA is less accurate for strong $V_\qs$ and, consequently, the simulations of the actual periodically driven system in Fig.~\ref{fig:6}(c) have a partially smoothed out step structure in $\Iav(\mu)$.

\subsection{Periodic driving of the lattice tilt}
\label{sec:holthaus}

Let us consider a second example of a periodically driven system that is well known from cold atom physics~\cite{eckardt_superfluid-insulator_2005}.
We consider a lattice with the tilt as in Eq.~\eqref{eq:HS}, but now we have time--independent {hopping} $v_\qs(t)=V_\qs$ and a periodically driven tilt {$\DeltaS(t) = \DeltaS \cos(\omega t)$ in Eq.~\eqref{eq:tilt}.}
As shown in Ref.~\onlinecite{eckardt_superfluid-insulator_2005}, for sufficiently large $\omega$, the system behaves as an effective time--independent model with no tilt, and the effective tunneling amplitude between sites equals 
\begin{equation}
    \label{eq:bessel}
    v_\qs^{\rm{eff}} = \VS {\cal J}_0(\DeltaS/\omega), 
\end{equation}
where ${\cal J}_0(\cdot)$ is a Bessel function of the first kind and order zero.
The relation holds for bosons~\cite{eckardt_superfluid-insulator_2005} and for fermions.
In effect, the tunneling is suppressed close to the zeros of the Bessel function {of the first kind}.
For interacting bosons in an optical lattice, it has been experimentally verified, that a transition occurs from the superfluid state in the absence of driving to a Mott insulator when tunnelings are effectively eliminated~\cite{lignier_dynamical_2007}.
Similar models for transport with $\NS=2$ have been considered in Refs.~\cite{kohler_charge_2004,chen_simple_2014,lacerda_quantum_2023,brenes_particle_2022}.

\begin{figure}[t!]
    \includegraphics[width=\columnwidth]{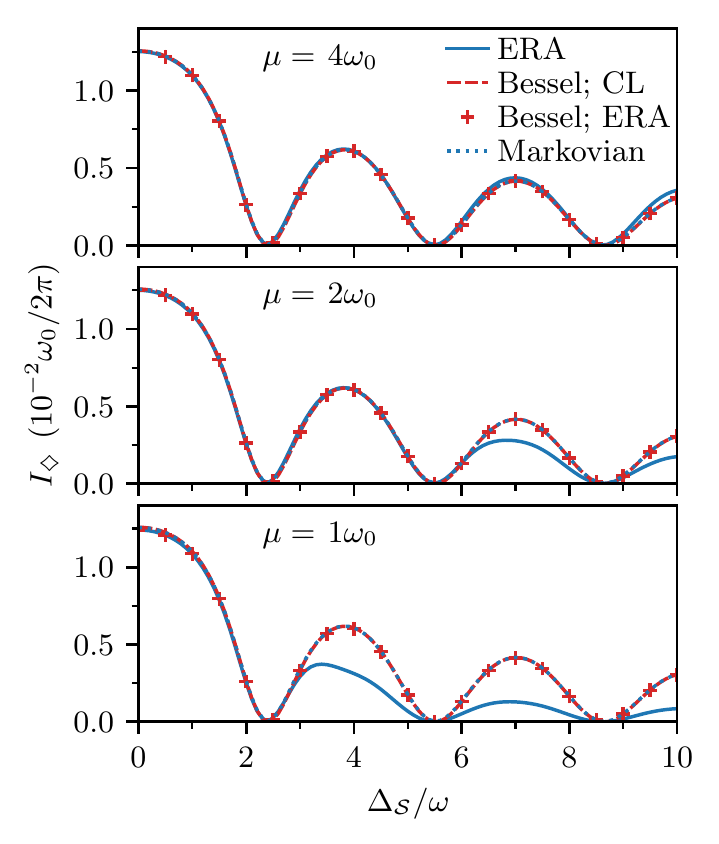}
    \caption{
    {\bf Periodically modulated tilt.}
    The current is shown versus the tilt. 
    We fix the driving frequency $\omega = \omega_0 / 20$ to be much larger than the hopping $\VS=\omega_0 / 200$, allowing one to approximate a time--dependent setup with a time--independent one with the hopping modulated by a Bessel formula according to Eq.~\eqref{eq:bessel}. The result of the latter is plotted directly in the continuum limit (dashed lines) and using ERA simulations (symbols).
    The approximation works exceptionally well for large enough bias and bandwidth (also captured by the Markovian limit, dotted lines). Reducing the bias $\mu$, which becomes comparable with the amplitude of oscillating tilt for large enough $\DeltaS$, leads
    to the qualitative breaking of the approximation for large tilts.
    The data are for $\NS=8$, $v_0 = \omega_0 / 20$, and the  bandwidth $\qw=4.0\omega_0$. 
    ERA simulations are for $\NW=512$ and $\gamma=2\Delta_F$, and checked for stability.
    }
    \label{fig:7}%
\end{figure}

Here, we shall consider a one--dimensional lattice with $N_\qs=8$ coupled to reservoirs. 
When the driving frequency is the largest scale, in Fig.~\ref{fig:7}(a), we can indeed see that the time-independent system approximation with hopping in the system given by Eq.~\eqref{eq:bessel} faithfully captures the behavior of the current.
We observe, however, a correlation between the amplitude of the total tilt and the bias $\mu$, i.e., an interplay between system dynamics and properties of the reservoirs.
Our ERA simulations of the periodically driven system, in Figs.~\ref{fig:7}(b) and \ref{fig:7}(c), illustrate that the approximation remains quantitatively valid when the amplitude of $\TotalS$ fits into the bias window set by $\mu$. 
With increasing $\Delta_\qs$ (that translates to $\TotalS$, which starts extending beyond the bias window), the current in a periodically driven system gets suppressed compared to the approximate time-independent prediction.
Notwithstanding, the zeros of the current coincide well with the zeros of the Bessel function in Eq.~\eqref{eq:bessel} also in that limit.

\section{Conclusions}
\label{sec:conclusions}

We benchmarked the application of ERA to simulate transport through a periodically driven system coupled to macroscopic reservoirs.
We focused on a tilted fermionic lattice with periodic driving.
Standard time--independent approximations of that model allow us to test proper convergence of the method in the corresponding limits.
We also study the properties of the setup in the parameter limits when the approximations can no longer be faithfully applied.

Our results exemplify potential traps in simulating transport properties using ERA--like approaches, in particular for time--dependent setups.
First, RWA mapping and the resulting effective shifts of the reservoir bands and bias window illustrates that discretization techniques promoting the bias window, like linear--logarithmic strategy, should be applied only with care.
A discretization strategy that distributes modes more evenly inside the whole reservoir band, like the one we use in this article, is less prone to misrepresentation of the reservoirs. 
Second, we introduce a stability criterion to properly tune the simulation parameters (the relaxation rates of the extended reservoirs).
It provides a model--agnostic tool to systematically avoid anomalous effects within ERA due to the interplay of the discretization of the continuum reservoirs and insufficient mode broadening.
Our results pave the way for faithful simulation of  transport in many--body, periodically driven quantum systems with tensor network and other techniques.

\begin{acknowledgments}
B.D. and G.W. contributed equally to this work. We gratefully acknowledge Polish high-performance computing infrastructure PLGrid (HPC Centers: ACK Cyfronet AGH) for providing computer facilities and support within computational Grant No.~PLG/2022/015613.
G.W. acknowledges the Fulbright Program and Michael Zwolak for hospitality during the Fulbright Junior Research Award at the National Institute of Standards and Technology.
This research has been supported by the National Science Centre (Poland) under Project No. 2019/35/B/ST2/00034 (B.D.), 2020/38/E/ST3/00150 (G.W. and M.M.R.) and under the OPUS call within the WEAVE program 2021/43/I/ST3/01142 (J.Z.).
The research has been supported by a grant from the Priority Research Area (DigiWorld) under the Strategic Programme Excellence Initiative at Jagiellonian University (J.Z., M.M.R.). 
\end{acknowledgments}

\begin{appendix}
\renewcommand{\theequation}{A\arabic{equation}}
\setcounter{equation}{0}
\renewcommand{\thefigure}{A\arabic{figure}}
\setcounter{figure}{0}
\section{Correlation matrix}
\label{sec:cm}
For a non--interacting Hamiltonian $H_\qs$, the evolution generated by Eq.~\eqref{eq:evolution_cr} preserves the Gaussianity of the density matrix.
For a particle number conserving $H(t)$, the latter is fully characterized by the correlation matrix
\begin{equation}
    \cm_{mn} = \tr [ \cid{n} \ci{m} \rho ],
\end{equation}
with $m,n\in\ql\qs\qr$.
The correlation matrix of a state evolved with Eq.~\eqref{eq:evolution_cr} follows a dynamic equation
\begin{equation}
    \label{eq:dotcm}
    \dot{\cm}(t) = - \im \left[\bH(t), \cm(t) \right] + \qd [ \cm(t) ],
\end{equation}
that can be efficiently integrated numerically.
Above, a single--particle Hamiltonian
\begin{equation}
H(t) = \sum_{m,n\in\ql\qs\qr} [\bH(t)]_{mn} \cid{m} \ci{n},
\end{equation}
and the dissipator in Eq.~\eqref{eq:dissipator} translates to~\cite{elenewski_communication_2017},
\begin{equation}
    \label{eq:CMdissipator}
    \qd [ \cm(t) ] = \Zop - \cm(t) \Gop - \Gop \cm(t),
\end{equation}
with matrices $\Zop = \sum_{k \in \ql \qr} \gamma_k^+ \proj{k}{k}$ and $\Gop = \frac{\gamma}{2} \sum_{k \in \ql \qr} \proj{k}{k}$. Here, we use notation where $\ket{k}$ is a column vector with value one for mode $k$ and zero for all other modes in $\ql \qs \qr$.

As we are interested in a Floquet state, we consider the correlation matrix evolution over a single cycle with period $\tau = 2 \pi / \omega$, which gives a map of the form
\begin{equation}
    \label{eq:Cmap}
    \cm(t_0 + \tau) = \Mop(\tau) \cm(t_0) \Mop^\dagger(\tau) + \Pop(\tau).
\end{equation}
In a steady state, $\cm(t_0 + \tau) = \cm(t_0)$, Eq.~\eqref{eq:Cmap} is a discrete Lyapunov equation~\cite{purkayastha_lyapunov_2022, landi_nonequilibrium_2022, wojtowicz_accumulative_2023} that allows finding the Floquet steady state numerically efficiently.
The same strategy was very recently taken in Ref.~\cite{brenes_particle_2022}.
Additionally, the steady state is unique if all eigenvalues  of $\Mop(\tau)$ have a magnitude smaller than one.
This condition is satisfied in all our examples.
However, analyzing the map in Eq.~\eqref{eq:Cmap} has an extra advantage, compared with a direct time evolution of some initial state, as it allows to directly probe for phenomena such as time crystals, which would require degenerate Floquet states~\cite{sacha_time_2018}.

The equations of motion for {the propagator} $\Mop(\tau)$ and {the source term} $\Pop(\tau)$ follow directly from Eq.~\eqref{eq:dotcm}, 
\begin{eqnarray}
    \dot{\Pop}(t) & = & - \im \left[\bH(t_0 + t), \Pop(t) \right] + \qd [ \Pop(t) ], \nonumber \\
    \dot{\Mop}(t) & = & \left[- \im \bH(t_0 + t) + \Gop \right] \Mop(t).
\end{eqnarray}
They can be efficiently numerically integrated over time $\tau$ with the two initial conditions, $\Mop(0)$ as an identity matrix and $\Pop(0)$ as a zero matrix.
We note that $\Pop(t)$ and $\Mop(t)$ also depend on the initial time $t_0$ that marks a conventional beginning of a single periodic cycle. 
We suppress it in the notation for simplicity.

\renewcommand{\theequation}{B\arabic{equation}}
\setcounter{equation}{0}
\section{Markovian limit}
\label{sec:markovian}

{The Markovian approximation in Eq.~\eqref{eq:markov_limit} for the infinite--bandwidth and infinite--bias limit follows from the normal Born--Markov master equation. One considers a single system mode of frequency $\omega_i$ connected to a fully-occupied reservoir (see Ref.~\onlinecite{elenewski_communication_2017} for extended discussion). The time correlation function of the reservoir with spectral function in Eq.~\eqref{eq:J1d} reads as
\begin{eqnarray}
    \mathcal{J}_+(t') & = & \frac{1}{2\pi} \int_{-\qw/2}^{\qw/2} \frac{8 v_0^2}{\qw^2} \sqrt{\qw^2 - 4 \omega^2} e^{\im \omega t'} d\omega  \\
    & = & \frac{4 v_0^2 \mathcal{J}_1(t' \qw /2) }{ t' \qw}, \nonumber
\end{eqnarray}
where $\mathcal{J}_1(\cdot)$ is the Bessel function of the first kind and order one.
The effective relaxation is
\begin{eqnarray}
    \gamma_\ql^+ &=& \int_0^\infty 2 \mathcal{J}_+(t') e^{-i \omega_i t'} dt'  \\
        &=& \frac{8 v_0^2}{\qw} \left( \sqrt{1 - 4 \omega_i^2 / \qw^2} + \im \omega_i / \qw \right) \nonumber
\end{eqnarray}
for system mode inside the bandwidth, $2 |\omega_i| < \qw$. 
Expanding to the leading order in the system frequency $\omega_i$ gives Eq.~\eqref{eq:gammaLR} up to corrections of order $\omega_i / \qw$. 
Note that, in reality, the system may have many frequencies but these all influence the relevant parameters in higher orders. 
The effective depletion rate $\gamma_\qr^-$ follows similarly.
}

\renewcommand{\theequation}{C\arabic{equation}}
\setcounter{equation}{0}
\section{Landauer formula}
\label{sec:landauer}

For a time--independent, non--interacting model, we can calculate the current flowing though the system directly in the continuum limit using non--equilibrium Green's functions~\cite{meir_landauer_1992, jauho_time-dependent_1994}.
We employ it for our approximate time--independent reference models, further corroborating proper convergence of ERA results to the continuum limit.
We collect the relevant equations here.

The retarded (advanced) Green's function for $\qs$ is
\begin{equation}
\bG^{r (a)}(\omega) = \frac{1}{\omega - \bHS - \bS^{r (a)}_\ql(\omega) - \bS^{r (a)}_\qr(\omega)}, 
\end{equation}
where the single--particle system Hamiltonian is
\begin{equation}
    H_\qs = \sum_{i,j\in\qs} [\bHS]_{ij} \cid{i} \ci{j}.
\end{equation}
The retarded (advanced) self--energies follow as
\begin{equation}
    \label{eq:SE}
    \bS^{r (a)}_\alpha(\omega) =  \int \frac{d \omega'}{2 \pi} \frac{\bJ_\alpha(\omega')}{\omega - \omega' \pm \im \eta},
\end{equation}
where $\bJ_\alpha(\omega)$ is the spectral function defining reservoir $\alpha$ and the limit of $\eta \to 0^+$ is taken at the end of the calculation. 
These quantities give the spectral densities  $\bGa^{\alpha}(\omega) = \im (\bS^r_{\alpha}(\omega) - \bS^a_{\alpha}(\omega)) = -2 \Im \bS^r_{\alpha}(\omega)$ 
(note that these are the spectral functions, but we retain both sets of terminology to correspond to other literature).
With this, the current is given by the Landauer formula, 
\begin{equation} 
    \label{eq:nonintCurrStandard}
    I = \int\frac{d\omega}{2\pi} \left( f^\ql (\omega) - f^\qr (\omega) \right) \tr \left[ \bGa^\ql \bG^r \bGa^\qr \bG^a \right],
\end{equation}
where $f^\alpha (\omega)$ is the Fermi-Dirac distribution in Eq.~\eqref{eq:fd}. 

Note that, in the RWA we employ in Sec.~\ref{sec:rwa}, the reservoir spectral functions in Eq.~\eqref{eq:J1d} get shifted, and the single non--zero element of Eq.~\eqref{eq:J1d} now reads as 
\begin{equation}
    J_\alpha (\omega) = \frac{8 v_0^2}{\qw^2} \sqrt{\qw^2 - 4 (\omega - \bar \omega_\alpha)^2},
\end{equation}
with $\bar \omega_\alpha = \pm \omega \TotalS / 2 \DeltaS$ for $\alpha = \ql (\qr)$, respectively.
The self--energies follow from Eq.~\eqref{eq:SE} as
\begin{equation}
    \Sigma^{r (a)}_\alpha(\omega) = \frac{2}{\omega - \bar \omega_\alpha \pm  \im  J_\alpha (\omega)},
\end{equation}
where again we only write the non-zero matrix element.
The integration interval in Eq.~\eqref{eq:nonintCurrStandard} is reduced to overlapping parts of shifted reservoir bandwidths where $|\omega - \bar \omega_\alpha| < \qw / 2$.

\end{appendix}

\end{document}